\documentstyle[12pt]{article}
\def\be{\begin{equation}}
\def\ee{\end{equation}}
\def\ba{\begin{eqnarray}}
\def\ea{\end{eqnarray}}

\def\bq{\begin{quote}}
\def\eq{\end{quote}}

\def\PRL{{\it Phys. Rev. Lett.} }

\def\PR{{\it Phys. Rev.} }
\def\MPL{{\it Mod. Phys. Lett.} }
\def\IJMP{{\it Int. J. Mod .Phys.} }

\newcommand{\beq}{\begin{equation}}
\newcommand{\eeq}{\end{equation}}
\newcommand{\beqa}{\begin{eqnarray}}
\newcommand{\eeqa}{\end{eqnarray}}

\begin{document}
\thispagestyle{empty}
\begin{flushright}
SU-ITP-99/43\\ hep-th/9909127\\ September 1999
\end{flushright}
\vspace*{1cm}
\begin{center}
{\Large \bf  Localized Gravity on Branes in anti--de Sitter Spaces}\\
\vspace*{1.5cm}
Edi Halyo\footnote{email: halyo@dormouse.stanford.edu} \\
\vspace*{0.2cm}
{\it Department of Physics}\\
{\it Stanford University}\\
{\it Stanford, CA 94305-4060}\\
\vspace{1.5cm}
ABSTRACT
\end{center}
We discuss the conditions under which  4D gravity is localized on
domain walls in 5D anti--de Sitter spaces. Our approach
is based on considering the limits in which the localized gravity
decouples. We find that gravity is localized if the wall is located a finite
distance from the boundary of the anti--de Sitter space and has a finite
tension. In addition, it has to be a $\delta$--function source of
gravity.

\vfill
\setcounter{page}{0}
\setcounter{footnote}{0}
\newpage

\section{Introduction}

The only consistent theory of quantum gravity is superstring theory which lives
in ten dimensions. Traditionally the extra six  dimensions are compactified on
a manifold of string or Planck size in order to render them unobservable. Four
dimensional gravity is obtained from ten dimensional gravity by dimensional
reduction on the small six manifold. Recently, two interesting
modifications to this scenario have been investigated. On one hand, the
existence of D--branes
in string theory with only super Yang--Mills (SYM) interactions on their
world--volume was discovered.
The SYM theory arises from the low energy physics of the open strings which
live on the branes.
Gravity does not propagate on D--branes but only in the bulk of space which
surrounds them.
On the other hand, it was noticed that the internal compact six dimensional
manifold does not have to be string or Planck scale if only gravity can
propagate in the bulk. In fact, present day limits on Newton's law can be
accomodated with two or more rather large dimensions of
micron to milimeter size\cite{nima}. In this case, the weakness of gravity
relative to the gauge interactions is a result of the large compact dimensions.

More recently, it was argued that there is no need to compactify extra
dimensions at all in order to get four dimensional gravity. In ref. \cite{rs1}
it was shown that gravity
can be localized on domain walls in anti--de Sitter ($AdS$) space.
In this manner, one can eliminate compactification
with all its well--known problems like moduli, stability etc. At first sight
this seems to contradict the well--known AdS--CFT duality which relates a bulk
gravitational theory to a boundary field theory. In particular on the boundary
of the $AdS$ space there is no gravity.
This work arose from trying to clarify this apparent contradiction.

In this letter, we find the necessary conditions for localization of gravity on
domain walls in $AdS$ space. In the following, we will use the terms brane or
domain wall interchangably. The branes in this language do not necessarily
refer to D--branes of string theory. We analyze the conditions under which the
four dimensional gravity decouples and
interpret the absence of these conditions as the necessary conditions for
localization of gravity. We find that
gravity is localized on branes in $AdS$ space if the brane is a finite distance
away from the boundary of the $AdS$ space and has a finite tension. It is also
crucial for the brane to contribute to
the stress--energy tensor as a $\delta$--function source as for domain walls in
supergravity.
These are the conditions that seem necessary for localizing gravity on the
brane world--volume. The reason is related to the nonnormalizable bulk graviton
modes in $AdS$ space. In pure $AdS$ space there is no gravity on the boundary
since the nonnormalizable graviton modes cannot be excited there. However, in
the presence of a brane and under the above conditions these modes are
eliminated. The new space with the brane at its boundary is not $AdS$ but
a truncated version of it which is symmetrical on both sides of the brane.
As a result, gravity can propagate on the boundary of the new space given by
the brane world--volume.  The above also applies to more general cases such as
intersections of branes in
 higher dimensional spaces\cite{savas,multi}. Recently, there have been a
number of papers investigating different aspects of this and related
scenarios\cite{nema}-\cite{etc}.

The letter is organized as follows. In the next section we review the solution
for the 3--brane in
$AdS_5$ and localization of gravity in four dimensions. In section 3 we show
that 4D gravity decouples in certain limits. The absence of these are
interpreted as the necessary conditions for localized gravity. Section 4
includes are conclusions and a discussion of our results.

\section{Localized Gravity on Branes}

We begin by reviewing the solution for a 3D domain wall  in $AdS_5$ located a
finite distance away from the boundary of $AdS_5$. In this configuration the
five dimensional graviton has
a massless normalizable zero mode on the world--volume of the
brane\cite{rs1},\cite{savas}. Therefore gravity is localized or trapped in four
dimensions.
Consider a $3D$ domain wall
in $AdS_5$ a finite distance from the boundary. We assume that the brane
couples only gravitationally to the $5D$ bulk theory.
The action specifying the dynamics of the brane-bulk system is
\be
S = \int_M d^5x \sqrt{g} \Bigl\{\frac{R}{2\kappa^2_5}
+ \Lambda \Bigr\} - \int_{\partial M} d^4x \sqrt{g} \sigma
\label{action}
\ee
where the bulk cosmological constant, $\Lambda$, is negative (since the bulk is
$AdS_5$),
the brane tension $\sigma$ is positive and $\kappa_5^2=8 \pi/M_5^3$ where $M_5$
is the fundamental mass scale of the bulk theory.

The equations of motion which follow from (3) are
\be
R^{\mu}{}_{\nu} - \frac12 \delta^{\mu}{}_{\nu} R = -\kappa_5^2
\sigma \delta(z-l) {\rm diag}(1,1,1,1,0) + \kappa_5^2
\Lambda  \delta^{\mu}{}_{\nu}
\label{ees}
\ee
where the stress-energy tensor is a combination of  the domain wall situated at
$z=l$
and the $AdS_5$ bulk terms. Note that the wall contribution is given by a
$\delta$--function in the $z$ direction which cuts all bulk gravitational modes
on the wall.
We are looking for solutions to the above equations
 which reflect the symmetries of the $3$-brane, i.e. $3+1$ dimensional Lorentz
symmetry.
The solution is given by (in Poincare coordinates, in terms of the shifted
coordinate $z^{\prime}=z+l$)\cite{rs1},\cite{savas}
\be
ds^2_5={L^2 \over {(|z|+l)^2}}(\eta_{\mu \nu}dx^{\mu}dx^{\nu}+dz^2)
\ee
Here $L$ is related to the bulk cosmological constant and $M_5$ by
$L^2={6 / {\kappa_5^2 \Lambda}}$
and $l$ is the distance of the brane to the boundary of $AdS_5$. This is
similar to but not quite $AdS_5$
due to the absolute value sign and the nonzero distance $l$. In fact, in this
metric the portion of $AdS_5$ from the brane to the boundary is deleted and the
bulk geometry is symmetric
on both sides of the brane due to the absolute value sign. The difference
between this space and $AdS_5$ are essential for localization of gravity on its
boundary, i.e. the brane world--volume.
The metric in eq. (3) is a solution of Einstein's
equations only if the tension of the wall is fine tuned to be $\sigma=6
\kappa_5^2 l^2/L^3$.

In order to show that gravity is localized on the brane we solve the equation
of motion for the linearized graviton in the above background
\be
\left[\frac{1}{2} \Box_4 + \left(-\frac{1}{2}
\nabla^2_z + V(z) \right) \right] \hat{h}(x,z) = 0,
\label{poteq}
\ee
where
\be
V(z) = \frac{3}{2L^2} \Omega(z)^2
- \frac{3}{2L} \Omega(z)  \delta(z).
\label{pot}
\ee
and $h=\Omega(z)^{-3/2} \hat h$ with
$\Omega(z)={L / {(|z|+l)}}$.

The massless four dimensional mode of the graviton is obtained by
setting $\hat{h}(x,z) = e^{i p x} {\psi}(z)$ and solving the effective Schr\"
odinger equation
\be
\left(-\frac{1}{2} \nabla^2_z + V(z)
\right) {\psi}_\lambda = \frac{1}{2} m^2_\lambda {\psi}_\lambda,
\ee
where $\lambda$ labels the eigenfunctions.
This potential has a repulsive piece which
goes to zero for $|z| \gg L$, and an
attractive $\delta$ function.
The $4D$ massless graviton
corresponds to a bound state with the
wavefunction
\be
\psi(z) =\sqrt{l^2 \over L^3} \Omega(z)^{3/2}.
\ee

In this case, the 4D gravitational action is given by
\be
S=M_5^3\int dz \Omega^3 \times \int d^4x \sqrt{g^{(4)}}R^{(4)}
\ee
Thus, the four dimensional graviton couples with the strength
\be
M_P^2=M_5^3 \int dz \Omega^3 = {M_5^3L^3 \over l^2}
\ee

The result for $M_P$ is identical to the case in which 5d gravity is
compactified on a circle of
size $L^3/l^2$. It can also be shown that gravity behaves as if it is four
dimensional down to
distances $\sim L^3/l^2$ whereas for shorter distances it becomes the higher
dimensional bulk gravity\cite{savas}.
We see that gravity behaves as if it is compactified on a circle whereas
in reality the fifth dimension is infinite.

As  long as $l \not =0$ i.e. the brane is not on the boundary of $AdS_5$ all
values of $l$ give equivalent but different solutions. This can be seen from
the tranformation $x^{\mu} \to \lambda x^{\mu},
z \to \lambda^{-1} z$ which is a broken symmetry (by the nonzero value of $l$)
of the solution in eq. (3). This transformation takes $l \to l/\lambda$
which corresponds to a new solution. Different values of $l$ are superselection
sectors
of the theory since the brane world--volume is infinite.
If we want we can conveniently choose $\lambda$ so that $l/\lambda=L$ in which
case the metric becomes that of ref. \cite{rs1}. (In the following we assume
that $l \not =L$ in order to clarify the role of the distance to the boundary
$l$.)
This symmetry is closely related to
the well--known scaling--radial translation equivalence of the AdS--CFT
duality. In that context,
it is known that a scale transformation on the boundary is equivalent to a
radial displacement
in the bulk. This is precisely the transformation we considered above.

\section{Decoupling Localized Gravity}

In this section we find the necessary conditions for localizing bulk gravity on
a lower
dimensional brane world--volume. We only consider 3--branes in bulk $AdS_5$
however our results can be easily generalized to higher dimensions.
The method we employ is to find out under what conditions or in which limits
the localized  4D gravity decouples.
We interpret the absence of these conditions as necessary conditions for
localizing gravity.
Without a brane in $AdS_5$ bulk there is no gravity on the boundary as is
well--known from the AdS--CFT duality which relates a gravitational bulk theory
to a field theory on the boundary.
This is due to the fact that there are nonnormalizable bulk gravitational modes
which cannot be excited on the boundary.
Localization of gravity on the brane is related to the fact that the presence
of the brane
eliminates these nonnormalizable modes.

We first consider the limit $l \to 0$. In this limit the brane approaches the
boundary and overlaps with it.
{}From the solution in eq. (3) we see that in this limit the space--time goes
to $AdS_5$ which has no
gravity on its boundary. This means that localized 4D gravity decouples in this
limit. We also see from eq. (9) that when $l \to 0$, $M_P$ diverges. This is
another indication that 4D gravity decouples in this limit. Thus, the first
condition for localization of gravity is the presence of a brane at a finite
distance from the boundary of $AdS_5$.

We can also see this from the bound state wavefunction in the transverse $z$
direction given by
$\psi(z) \sim  l \Omega(z)^{3/2}$.
It is easy to see that when $l \not=0$, the bound state wavefunction $\psi(z)$
is normalizable and gravity is localized. On the other hand, when $l=0$,
$\psi(z)$ is not normalizable. Therefore there is no bound state in the
transverse direction and gravity cannot be localized on the brane. The full
geometry of the brane in $AdS_5$ is a truncated version of $AdS_5$ which
eliminates the nonnormalizable bulk graviton
modes.

The second limit we consider is $l$ finite and fixed but $\sigma \to 0$, i.e.
the limit in which the tension of the brane vanishes. This is a nontrivial
limit because in order for eq. (3) to be a solution to Einstein's equations
we need $\sigma=6/\kappa_5^2 L$ or using the definition of $L$, $\sigma=\sqrt{6
\Lambda}/\kappa_5$. Because of this relation we cannot take the $\sigma \to 0$
limit and keep
the bulk curved. In order to be able to take $\sigma \to 0$ and keep $\Lambda$
fixed we consider a system of two branes at a distance  with positive tensions
$\sigma_1$ and $\sigma_2$\cite{nema}. The two branes can be at a finite
distance only if $6 \Lambda>\kappa_5^2 \sigma_{1,2}^2$. Note that due to the
this condition the brane world--volumes are $AdS_4$ slices of the 5D metric
with negative overall energy density.
The metric which corresponds to this case is given by\cite{nema}
\be
ds^2_5 = a^2(w)
\Bigl(dx^2 + e^{2{\cal H} x}(-dt^2 + dy^2 + dz^2)
\Bigr) + dw^2
\label{metansads}
\ee
with the warp factor
\be
a = \cosh(\sqrt{\frac{\kappa^2_5\Lambda}{6}}w)
- \frac{\kappa_5 \sigma}{\sqrt{6\Lambda}}
\sinh(\sqrt{\frac{\kappa^2_5 \Lambda}{6}} |w|)
\label{scalefac}
\ee
Here ${\cal H}=iH$ where
\be
H^2  = \kappa^2_5 \frac{{\kappa_5^2\sigma^2 - 6\Lambda}}{36}
\label{Hubble}
\ee
Due to the presence of the two branes at $w=0$ and $w=w_c$ and the absolute
value sign
in the metric the identifications $w \sim -w$ and $w \sim w+w_c$ have to be
made in (10). This is precisely what would be obtained by compactifying the
transverse direction on a circle of size $w_c$ and orbifolding by $Z_2$.
This metric is equivalent to the one in eq. (3) when $H=0$ with the
redefinition of the
transverse coordinate $Lexp(|w|/L)=|z|+l$. Note that the condition $H=0$ can be
written as
$\sigma=6/\kappa_5^2L$ using the definition of $L$.
It can be shown that the
branes are separated by a distance $w_c$ given by
\be
\tanh(\sqrt{\frac{\kappa^2_5 \Lambda}{6}} w_c) =
\frac{\kappa_5 \sqrt{6\Lambda}(\sigma_1 +  \sigma_2)}{6 \Lambda
+ \kappa^2_5 \sigma_1 \sigma_2}
\label{distconthreet}
\ee

We would like to use this configuration with two branes to reach a
configuration with only one brane which does not satisfy the relation
$\sigma=6/\kappa_5^2 L$. In order to get rid of the second brane we take the
distance between the branes to infinity. We see that
$w_c \rightarrow \infty$
requires
\be
(\sqrt{6\Lambda}-\kappa_5 \sigma_1) (\sqrt{6\Lambda}-\kappa_5 \sigma_2)=0
\ee
By choosing the tension of the second brane to satisfy $6 \Lambda=\kappa_5^2
\sigma_2^2$ we send it to infinity. As a result, we are left with only one
brane with positive tension $\sigma_1$ in $AdS_5$, but now there is no
correlation between $\sigma_1$ and $\Lambda$.
This is precisely the situation we wanted to examine for the limit $\sigma_1
\to 0$.
In this picture, the second brane with a tension fixed by $\Lambda$ plays a
role similar to a regulator. Now consider a very large but finite distance
between the two branes. Then using
\be
M_P^2 = M_5^3
\int^1_0 d\vartheta w_c a^2(w_c \vartheta)
\label{planckmass}
\ee
where $a(w)$ is the warp factor we find that
\be
M_P^2={6 M_5^6 \over \sigma_1}
\ee
We see that in the limit $\sigma_1 \to 0$, $M_P$ diverges and 4D gravity
decouples.

The third condition for localizing gravity on branes is the requirement that
the brane contribute
as a $\delta$-function to the stress--energy tensor which is the source of
gravity as in eq. (2).
In supergravity in $AdS$ space with a domain wall this is indeed the case.
Consider for example the effective low--energy bosonic action (in D space--time
dimensions)
\be
{\cal L}=\sqrt{g}\left({R \over {2 \kappa_D^2}}-{1 \over 2}(\partial \phi)^2+
\Lambda e^{-a \phi}\right)
\ee
which appears in dimensionally reduced supergravity where
\be
a^2=\Delta+{2(D-1) \over {D-2}}
\ee
is a constant. Note that this is similar to eq.(1) without the brane source
term but with a dilaton coupling. For $a=0$
there is the well--known solution which is anti--de Sitter space. For $a \not
=0$ however, the solution is a
$D-2$ dimensional domain wall\cite{pope}
\be
ds^2=H^{4/\Delta(D-2)} \eta_{\mu \nu} dx^{\mu} dx^{\nu}+H^{4(D-1)/\Delta(D-2)}
dy^2
\ee
situated at $y=0$ with $H(y)=1+k|y|$. (here $H$ is $\Omega$ of the previous
solution
with $k=1/L$.)
The curvature is smooth everywhere except at $y=0$ where it has a
$\delta$--function singularity. This is due to the brane tension which must be
included in eq. (17) for consistency.
This $\delta$-function
source cuts all graviton modes on both sides of the brane, in particular it
cuts the nonnormalizable modes of the bulk graviton which allow gravity to be
localized on the brane.

Finally, there can be domain wall configurations which look like eq. (3) but
with a power $p$
of $\Omega$ different than two as in eq. (19)\cite{edi1}. Such configurations
appear in theories obtained by dimensional reduction of higher dimensional
supergravity. Then from eq. (9)
\be
M_P^2=M_5^3 \int dz \Omega^{3p}
\ee
Thus $M_P$ is finite and there is 4D gravity only if $p<-1/3$. The same bound
also guarantees that the bound state wavefunction in eq. (7) is normalizable.

\section{Conclusions and Discussion}

In this letter, we found the necessary conditions for localizing gravity on 3D
branes (or domain walls) in $AdS_5$. Gravity is localized on the brane if
3D brane is located a finite distance away from the boundary of $AdS_5$.
The four dimensional Planck scale is inversely proportional to this distance.
As a result, the smaller the distance to the boundary the weaker is 4D gravity.
When the brane is on the boundary, 4D gravity decouples as is expected from the
well--known AdS--CFT duality. Another condition for localizing gravity is
a finite brane tension. Since for these solutions, the brane tension is
proportional to the cosmological constant this cannot be examined in a trivial
manner. Thus, we started from two positive tension branes in $AdS_5$ and sent
one of them to infinity by fixing its tension. Then we found that 4D Planck
scale is inversely proportional to the tension of the remaining brane. As a
result, if the tension of the brane goes to zero 4D gravity decouples. The
third condition for localization
of gravity is that the contribution of the brane to the stress--energy tensor
must be a $\delta$--function. This is indeed the case for domain walls in
supergravity. The $\delta$--function cuts all momentum modes of the bulk
graviton
including the ones which are nonnormalizable in full $AdS_5$ without the brane.
We stress that the geometry with the brane in the bulk of $AdS_5$ is a modified
space where the part of $AdS_5$ from the brane to the boundary is deleted
and the space is symmetric on both sides of the brane. Since the bulk graviton
modes are now normalizable on the brane they can be excited and this gives rise
to 4D gravity. This can also be seen by inspecting the normalizable graviton
wavefunction in the fifth dimension transverse to the brane.

In refs. \cite{rs1},\cite{savas}, it was shown that gravity is localized on the
brane by solving
the linearized Einstein equations in the background of the brane in $AdS_5$.
One finds that there is a graviton zero mode in 4D which is localized around
the brane. This analysis is clearly perturbative since it considers only weak
fields. It is not clear whether the same result can be obtained for strong
gravity. In particular, it is not clear how to think about black holes either
in the bulk or on the brane in this context. For example, if there is a 5D
black hole inthe bulk, it is not clear how to reduce it to 4D or whether it
will look like a 4D black hole. Perturbative 4D gravity looks like 5D bulk
gravity compactified
on a circle of size $L^3/l^2$; however it does not seem that the same is true
for 5D black holes. Similarly, it is not clear how 4D black holes which must
exist look like in the bulk.

Another issue is whether localization of gravity which was shown in
(super)gravity can be realized in string theory. This has been the subject of
refs. \cite{strings} to some degree. In string theory a possible setup would be
IIB string theory on $AdS_5 \times S^5$ with a D3 brane in the bulk. The part
of $AdS_5$ between the brane and the boundary can be deleted by an orbifold
about the brane position. Naively, the appearance of gravity on the D3 brane
world--volume is very surprising since the D3 brane world--volume theory is
supersymmetric Yang--Mills theory at low energy. If gravity can be localized in
string theory it would be important to understand whether there is a
bulk/boundary duality as in the AdS--CFT case.
In the AdS--CFT duality the dimension transverse to the boundary is
holographic, i.e. it is related to the energy scale of the boundary theory. It
is not clear that
this will continue to be the case when gravity is localized on the brane.

On the more phenomenological side it is not clear how inflation can be
realized. Prevoiusly considered mechanisms for inflation on branes
such as brane inflation, asymmetric inflation and D--term inflation\cite{inf}
cannot be
easily used in this context. This is either due to the absence of the scalar
(parametrizing the distance between branes) and radion
fields or due to the absence of an anomalous $U(1)$ symmetry on the brane.
 In \cite{nema,cos} it was shown that if there is an overall nonzero vacuum
energy on the brane it will inflate. This requires a time dependent negative
cosmological constant in the bulk, e.g. a bulk scalar with a negative vacuum
energy and a potential suitable for inflation. Such scalars seem to exist in
gauged supergravity and may indeed lead to inflation on the brane\cite{edi2}.
It is important to see whether these scalars have potentials which can satisfy
all the requirements of acceptable inflation.

\section{Acknowledgements}

We would like to thank Raphael Bousso, Steve Shenker and Lenny Susskind and
especially Nemanja Kaloper for useful discussions.

%
\def\IJMP #1 #2 #3 {Int. J. Mod. Phys. A {\bf#1},\ #2 (#3)}
\def\MPL #1 #2 #3 {Mod. Phys. Lett. A {\bf#1},\ #2 (#3)}
\def\NPB #1 #2 #3 {Nucl. Phys. {\bf#1},\ #2 (#3)}
\def\PLBold #1 #2 #3 {Phys. Lett. {\bf#1},\ #2 (#3)}
\def\PLB #1 #2 #3 {Phys. Lett. B {\bf#1},\ #2 (#3)}
\def\PR #1 #2 #3 {Phys. Rep. {\bf#1},\ #2 (#3)}
\def\PRD #1 #2 #3 {Phys. Rev. D {\bf#1},\ #2 (#3)}
\def\PRL #1 #2 #3 {Phys. Rev. Lett. {\bf#1},\ #2 (#3)}
\def\PTT #1 #2 #3 {Prog. Theor. Phys. {\bf#1},\ #2 (#3)}
\def\RMP #1 #2 #3 {Rev. Mod. Phys. {\bf#1},\ #2 (#3)}
\def\ZPC #1 #2 #3 {Z. Phys. C {\bf#1},\ #2 (#3)}


\begin{thebibliography}{99}

\bibitem{nima}
I. Antoniadis, Phys. Lett. {\bf B246} (1990) 377; J. Lykken,
 Phys. Rev. {\bf D54} (1996) 3693, hep-th/9603133;
N. Arkani-Hamed, S. Dimopoulos and G. Dvali, \PLB B429 263 1998 , and
\PRD D59 086004 1999 ; I. Antoniadis, N. Arkani-Hamed, S. Dimopoulos and
G. Dvali,  \PLB B436 257 1998 .
%
\bibitem{rs1}
L. Randall and R. Sundrum, hep-ph/9905221;
L. Randall and R. Sundrum, hep-th/9906064.
%
\bibitem{savas}
Arkani-Hamed, S. Dimopoulos, G. Dvali, and N. Kaloper, hep-th/9907209.
%
\bibitem{multi}
C. Csaki and Y. Shirman, hep-th/9908186;
A.E. Nelson, hep-th/9909001.
%
\bibitem{nema}
N. Kaloper, hep-th/9905210.
%
\bibitem{strings}
H. Verlinde, hep-th/9906182;
A. Brandhuber and K. Sfetsos, hep-th/9908116.
%
\bibitem{cos}
P. Binetruy, C. Deffayet and D. Langlois, hep-th/9905012;
T. Nihei, hep-ph/9905487;
C. Csaki, M. Graesser, C. Kolda, and J. Terning, hep-ph/9906513;
J.M. Cline, C. Grojean and G. Servant, hep-ph/9906523;
H. B. Kim and H. D. Kim, hep-th9909053:
U. Ellwanger, hep-th/9909103.
%
\bibitem{sugra}
A. Kehagias, hep-th/9906204;
Klaus Behrndt and Mirjam Cvetic, hep-th/9909058.
%
\bibitem{etc}
W.D. Goldberger and M.B. Wise, hep-ph/9907447; hep-ph/9907218;
J. Lykken and L. Randall, hep-th/9908076;
I. Oda, hep-th/9908104; 9909048;
K.R. Dienes, E. Dudas, and T. Ghergetta, hep-ph/9908530;
H. Davoudiasl, J. L. Hewett and T. G. Rizzo, hep-ph/9909225.
%
\bibitem{pope}
H. Lu, C. Pope and P. Townsend, hep-th/9607164.
%
\bibitem{edi1}
E. Halyo, to appear.
%
\bibitem{inf}
 D.H. Lyth,  Phys. Lett. {\bf B448} (1999) 191, hep-ph/9810320;
 N. Kaloper and A. Linde, hep-th/9811141;
N. Arkani-Hamed, S. Dimopoulos, N. Kaloper and J. March-Russell,
hep-ph/9903224;
E. Halyo, Phys. Lett. {\bf B454} (1999) 223; hep-ph/9901302;
E. Halyo, hep-ph/9905244; hep-ph/9907223.
\bibitem{edi2}
E. Halyo, work in progress.

\end{thebibliography}
\end{document}